\documentclass[lettersize,letterpaper]{IEEEtran}
\usepackage{amsmath,amsfonts}
\usepackage{algorithmic}
\usepackage{algorithm}
\usepackage{array}
\usepackage{textcomp}
\usepackage{stfloats}
\usepackage{url}
\usepackage{verbatim}
\usepackage{graphicx}
\usepackage{cite}
\usepackage{bm}
\usepackage{subcaption}
\usepackage{booktabs} % For better table lines
\usepackage{multirow}
\usepackage{amsthm} % 用于自定义定理和备注样式
\usepackage{hyperref} % 提供超链接功能和可以自定义引用样式
\usepackage{cleveref} % 提高引用功能，可以和hyperref一起使用
\usepackage{makecell}
\usepackage{amssymb}
\begin{document}
	
	\title{LLM-enabled Antenna Partitioning and Beamforming Optimization for Segmented Pinching Antenna assisted ISAC Systems}
	
	\author{Qian Gao,~\IEEEmembership{Graduate Student Member,~IEEE,} 
		Ruikang Zhong,~\IEEEmembership{Member,~IEEE,} \\
		Hyundong Shin,~\IEEEmembership{Fellow,~IEEE,}
		Yuanwei Liu,~\IEEEmembership{Fellow,~IEEE}

		\thanks{Qian Gao is with the School of Electronic Engineering and Computer Science, Queen Mary University of London, London, E1 4NS, U.K. (e-mail: q.gao@qmul.ac.uk).
		
		Ruikang Zhong is with the  School of Information and Communication Engineering, Xi’an Jiaotong University, Xi’an 701149, P.R. China. (e-mail: ruikang.zhong@xjtu.edu.cn)
		
		Hyundong Shin is with the Department of Electronics and Information
		Convergence Engineering, Kyung Hee University, 1732 Deogyeong-daero,
		Giheung-gu, Yongin-si, Gyeonggi-do 17104, Republic of Korea. (e-mail:
		hshin@khu.ac.kr)
		
		Yuanwei Liu  is with
		the Department of Electrical and Electronic Engineering, The University
		of Hong Kong, Hong Kong. (e-mail: yuanwei@hku.hk).
	
}
	}

	\maketitle

\begin{abstract}

Integrated sensing and communication (ISAC) requires spatial architectures that can flexibly balance data transmission and environment sensing. Segmented pinching antenna-assisted ISAC provides such flexibility by allowing different waveguide segments to be dynamically configured for transmission and reception. However, its design involves the joint optimization of antenna deployment, segment partitioning, and beamforming under coupled communication and sensing constraints, which becomes particularly challenging when the numbers of communication users and sensing targets vary across scenarios. To endow the system with stronger adaptability to changing user and target configurations, we propose a general learning framework for segmented pinching antenna-assisted ISAC systems. Specifically, a channel state information (CSI)-induced self-graph is constructed to produce permutation-invariant representations of user-target interactions, and the resulting features are processed by a large language model (LLM) backbone with two task-specific heads for jointly predicting antenna deployment, segment partitioning, and ISAC beamforming. In addition, a user count transfer mechanism is developed to examine whether the learned deployment policy is site-specific and reusable under changed user configurations. Simulation results show that the proposed framework achieves higher communication rates while maintaining reliable sensing accuracy. Moreover, the learned deployment policy remains highly stable when transferring to other user counts, which reduces the training cost from full model retraining to beamforming head adaption.

\end{abstract}


\section{Introduction}

Integrated sensing and communication (ISAC) \cite{ISAC6G} is regarded as a key technology for future sixth-generation (6G) wireless networks. It enables communication and sensing functionalities to share spectrum, hardware, and signal processing resources. By integrating these two functionalities, ISAC can improve spectral efficiency, hardware utilization, and environmental awareness, and thus support emerging applications such as autonomous driving, intelligent transportation, and smart manufacturing \cite{uav,driving,app}. However, the communication and sensing tasks are inherently coupled in the spatial domain, where limited spatial degrees of freedom (DoF)~\cite{doa} need to be carefully allocated to simultaneously guarantee communication throughput and sensing accuracy.

The antenna array architecture plays a fundamental role in determining the achievable spatial DoF of ISAC systems. In line-of-sight (LoS) dominated scenarios, conventional fixed arrays often have limited reconfigurability and may not fully exploit spatial resources. To improve spatial flexibility, several new antenna architectures have recently been studied, including intelligent reflecting surfaces \cite{ris}, movable antennas \cite{move}, and fluid antennas \cite{fuild}. Although these architectures provide additional spatial adaptability, they usually introduce non-negligible hardware complexity or mechanical control overhead. Recently, segmented waveguide-enabled pinching antenna (SWAN) architectures \cite{seg} have been proposed as a promising alternative. By dividing the waveguide into multiple independently controlled segments and employing pinching antennas as reconfigurable radiating elements, SWAN offers flexible and cost-effective LoS link establishment with high structural freedom \cite{segpinch}. In addition, each segmented waveguide together with the pinching antennas mounted on it \cite{PA} can be configured for transmission or reception, which provides new opportunities for spatial resource allocation in ISAC systems.

Based on array architectures, dynamic antenna selection and partitioning have been widely investigated to improve array efficiency and reduce hardware cost. For example, authors in\cite{selection1} proposed movable subarray selection and adjustable antenna selection schemes for extremely large-scale ISAC arrays, while \cite{selection2} developed heuristic and convolution neural network (CNN) based subarray selection methods to combat beam squint in terahertz ISAC. Joint transmit/receive antenna selection was further considered in \cite{selection3} to improve sensing capability in multiple-input multiple-output (MIMO) ISAC systems. However, antenna partitioning in SWAN is fundamentally different from that in conventional arrays. In conventional architectures, antenna selection is typically performed at the element level through RF switching. In contrast, in pinching antenna systems based on dielectric waveguides~\cite{PA1Ddl, PA1Dul, PA2D}, antenna behavior is coupled through the waveguide feed, and the segmented structure introduces an additional hierarchy in the design space. Specifically, different segments can be assigned to transmission or reception, while the antenna deployment over the segmented waveguides jointly determines the effective array geometry and beamforming capability. Therefore, SWAN-ISAC leads to a hierarchical structural optimization problem rather than a simple element-wise antenna selection problem.

The optimization of SWAN-ISAC is challenging for three reasons. Firstly, antenna deployment, segment-wise transmit/receive partitioning, and beamforming are strongly coupled, leading to a mixed discrete-continuous optimization problem. Secondly, in practical ISAC scenarios, the numbers of communication users and sensing targets may vary over time, which makes fixed-input optimization and learning methods difficult to generalize. Thirdly, communication utility and sensing accuracy must be optimized jointly, since sensing-oriented constraints directly affect feasible beamforming designs. These factors make conventional model-based optimization \cite{model1,model2} difficult to scale, especially for segmented pinching antenna-assisted ISAC systems with dynamic user configurations.

Learning-based methods provide a viable alternative for high-dimensional coupled design problems. Existing deep learning and reinforcement learning approaches \cite{dl1,dl2,dl3} have been applied to wireless resource allocation and beamforming tasks, but most of them assume fixed system dimensions and fixed user configurations. As a result, their applicability to segmented pinching antenna-assisted ISAC remains limited. Graph neural networks (GNNs)\cite{GCN} are appealing in this context because they naturally model interactions among users and targets and can produce permutation-invariant representations \cite{size}. By representing communication users and sensing targets as graph nodes, GNN-based models can better capture scenario-dependent relations than standard ordered-input neural networks. Nevertheless, existing GNN-based methods\cite{graphpinch1, graphpinch2} mainly focus on feature extraction or beamforming under fixed system structures, and they do not directly address the joint design of deployment, partitioning, and beamforming for SWAN-ISAC.

Recently, large language models (LLMs) \cite{llm1,llm2,llm3} have shown strong sequence modeling and generalization capability beyond natural language processing. Their expressive modeling ability makes them attractive for complex wireless design problems with heterogeneous inputs and coupled decision variables. However, directly applying an LLM to physical-layer optimization is nontrivial, since wireless inputs are structured numerical signals rather than text, and the learned model must remain compatible with system constraints and physics-based objectives. This motivates the development of structure-aware LLM-based learning architectures for segmented pinching antenna-assisted ISAC systems.

In this paper, we propose a general learning framework for segmented pinching antenna-assisted ISAC systems. The proposed framework combines a channel state information (CSI)-induced self-graph representation with an LLM backbone using low-rank adaptation (LoRA)\cite{lora}. Specifically, a CSI-induced self-graph neural network (SGNN) is first constructed to capture the interaction structure among communication users and sensing targets through CSI similarity, thereby producing permutation-invariant scenario representations for variable user settings. The graph-enhanced representation is then processed by an LLM backbone, followed by two task-specific output heads for antenna deployment and beamforming prediction, respectively. In this way, the proposed framework jointly predicts antenna deployment, segment partitioning, and communication/sensing beamforming variables in a unified manner. Furthermore, the proposed design allows us to investigate whether the learned deployment policy is site-specific and reusable under changed user configurations, while adapting only a lightweight beamforming head.

The main contributions of this paper are summarized as follows:
\begin{itemize}
	\item We formulate a joint design problem for SWAN-ISAC systems, where antenna deployment, segment-wise transmit/receive partitioning, and beamforming are jointly optimized under communication-rate and sensing-accuracy requirements. The formulation explicitly captures the hierarchical structural flexibility of the SWAN architecture.
	
	\item We propose a general learning framework that integrates a CSI-induced SGNN and an LLM backbone with LoRA. The SGNN provides permutation-invariant CSI representations for variable user settings, while two task-specific output heads are used to jointly predict deployment, and beamforming variables.
	
	\item We further investigate user-count transfer in SWAN-ISAC systems and show that the proposed framework can preserve a stable deployment policy across changed user configurations while significantly reducing the adaptation burden by focusing retraining on the beamforming head. Simulation results demonstrate a favorable tradeoff among communication performance, sensing accuracy, and transferability.
\end{itemize}

\section{System Model and Problem Formulation}\label{sec:system}

\begin{figure}[t!]
	\centering
	\captionsetup{justification=raggedright,singlelinecheck=false}
	\includegraphics[width=0.5\textwidth]{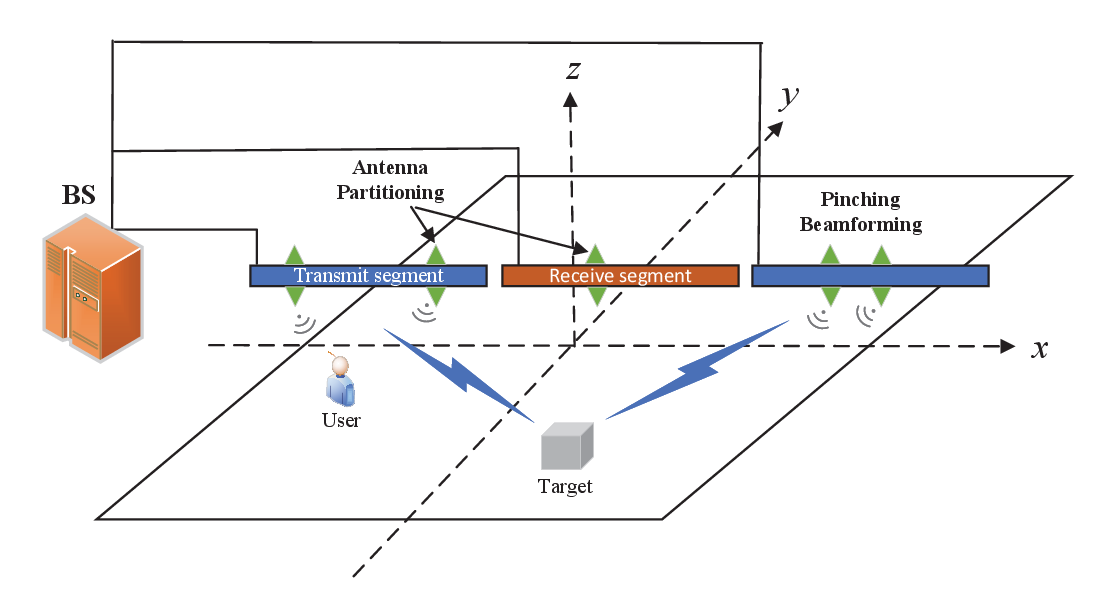}
	\caption{Illustration of the SWAN-ISAC system..}
	\label{Fig.1}
\end{figure}

We consider a SWAN-ISAC system, where a base station (BS) is equipped with a segmented dielectric-waveguide structure consisting of $M$ independently controllable segments and $N$ pinching antennas (PAs) to simultaneously serve $K_c$ communication users and sense $K_s$ targets, as illustrated in Fig.~\ref{Fig.1}. The dielectric waveguide is deployed at the BS side, and the PAs are placed along the waveguide axis to form a reconfigurable aperture. Different from a conventional fixed array, the segmented waveguide allows each segment to be individually configured for transmission or reception, such that different portions of the aperture can be flexibly allocated to downlink communication and sensing echo acquisition. As a result, the segment-wise transmit/receive assignment and the PA deployment over the segmented waveguide are tightly coupled, and together determine the effective array geometry as well as the communication and sensing performance.

For notational simplicity, we focus on one channel realization or one spatial scenario, and omit the time-slot index unless otherwise needed. The BS reference coordinates are denoted by $(x_{\mathrm{BS}},y_{\mathrm{BS}},z_{\mathrm{BS}})$, and all PAs are deployed along the $y$-axis over a waveguide length $L$. The position of the $n$th PA is written as
\begin{equation}
	\bm{\psi}_n = (x_{\mathrm{BS}}, y_n, z_{\mathrm{BS}}), \quad n\in\{1,\ldots,N\},
\end{equation}
where $y_n \in [0,L]$ denotes the deployment coordinate along the waveguide axis. The position vectors of the $k_c$th communication user and the $k_s$th sensing target are denoted by $\mathbf{p}_{k_c}\in\mathbb{R}^{3}$ and $\mathbf{p}_{k_s}\in\mathbb{R}^{3}$, respectively.

\subsection{Segment-Wise Transmit/Receive Partition}\label{subsec:partition}

Due to the segmented structure of SWAN, we first design the segment-wise transmit/receive (Tx/Rx) partition for the hierarchical structure. Let
\begin{equation}
	\chi_m \in \{0,1\}, \quad m\in\{1,\ldots,M\},
\end{equation}
denote the operating mode of the $m$th segment, where $\chi_m=1$ and $\chi_m=0$ indicate that the segment is configured for transmission and reception, respectively. Accordingly, the transmit-segment set and receive-segment set are defined as
\begin{align}
	\mathcal{M}_{\mathrm{t}} &\triangleq \{m \,|\, \chi_m=1\},\notag \\ 
	\mathcal{M}_{\mathrm{r}} &\triangleq \{m \,|\, \chi_m=0\},
\end{align}
with $M_{\mathrm{t}} = |\mathcal{M}_{\mathrm{t}}|$ and $M_{\mathrm{r}} = |\mathcal{M}_{\mathrm{r}}|$.

To support multiuser communication while reserving a nonzero receiving aperture for sensing echo acquisition, the segment partition should satisfy
\begin{equation}
	K_c +K_s\le M_{\mathrm{t}} \le M-1,
\end{equation}
where the segment-wise partition determines which parts of the segmented aperture contribute to downlink transmission and which parts are reserved for sensing reception.

\subsection{Antenna Deployment over Segmented Waveguides}\label{subsec:deployment}

Then, we design the antenna deployment over the segmented waveguides. Let
\begin{equation}
	\mathbf{y} = [y_1,\ldots,y_N]^T
\end{equation}
denote the deployment vector of the $N$ PAs along the waveguide axis. Since the waveguide length is $L$ and the structure is divided into $M$ equal segments, the $m$th segment corresponds to the interval
\begin{equation}
	\mathcal{I}_m = \left[\frac{(m-1)L}{M}, \frac{mL}{M}\right), \quad m=1,\ldots,M-1,
\end{equation}
and
\begin{equation}
	\mathcal{I}_M = \left[\frac{(M-1)L}{M}, L\right].
\end{equation}
Therefore, the segment index associated with antenna $n$ is determined by the interval that contains $y_n$.

To ensure physical feasibility, the antenna deployment should satisfy the range constraint
\begin{equation}
	0 \le y_n \le L, \quad \forall n,
\end{equation}
and the minimum-spacing constraint
\begin{equation}
	|y_n-y_{n'}| \ge d_{\min}, \quad \forall n\neq n',
\end{equation}
where $d_{\min}$ is the minimum allowable inter-antenna spacing, typically chosen as half wavelength. In addition, the deployment should remain compatible with the segmented hardware structure, so that the resulting effective aperture can be mapped to the segment-wise Tx/Rx partition.

Compared with conventional antenna selection, the deployment vector $\mathbf{y}$ directly determines the effective array geometry of the SWAN system. Therefore, the spatial design problem in SWAN-ISAC is not only a transmit/receive partitioning problem, but also a deployment optimization problem.

\subsection{Near-Field Signal and Channel Model}\label{subsec:signal_channel}

We consider a near-field spherical-wave channel model \cite{nearfield}. For a node located at $\mathbf{p}\in\mathbb{R}^{3}$, the channel from the deployed PAs to this node is expressed as
\begin{equation}
	\mathbf{h}(\mathbf{p})
	=
	\left[
	\frac{\alpha e^{-j\frac{2\pi}{\lambda}\|\mathbf{p}-\bm{\psi}_1\|}}{\|\mathbf{p}-\bm{\psi}_1\|},
	\ldots,
	\frac{\alpha e^{-j\frac{2\pi}{\lambda}\|\mathbf{p}-\bm{\psi}_N\|}}{\|\mathbf{p}-\bm{\psi}_N\|}
	\right]^T \in \mathbb{C}^{N\times 1},
\end{equation}
where $\alpha$ denotes the path-gain coefficient and $\lambda$ is the carrier wavelength. Accordingly, the channels toward the $k_c$th communication user and the $k_s$th sensing target are denoted by $\mathbf{h}_{k_c}\triangleq \mathbf{h}(\mathbf{p}_{k_c})$ and $\mathbf{h}_{k_s}\triangleq \mathbf{h}(\mathbf{p}_{k_s})$, respectively.

Let $\mathbf{w}_{k_c}\in\mathbb{C}^{N\times 1}$ denote the beamforming vector for the $k_c$th communication user and let $\mathbf{f}_{k_s}\in\mathbb{C}^{N\times 1}$ denote the probing beamforming vector for the $k_s$th sensing target. To reflect the segment-wise partition, the beamforming vectors are effectively constrained by the transmit-segment configuration, such that only antennas belonging to transmit-mode segments contribute to the transmitted signal. Let $\rho_c\in[0,1]$ and $\rho_s\in[0,1]$ denote the communication and sensing power-splitting factors, respectively, satisfying
\begin{equation}
	\rho_c+\rho_s \le 1.
\end{equation}
Then the transmitted baseband signal can be written as
\begin{equation}
	\mathbf{x}
	=
	\sqrt{\rho_c}\sum_{k_c=1}^{K_c}\mathbf{w}_{k_c} s_{k_c}
	+
	\sqrt{\rho_s}\sum_{k_s=1}^{K_s}\mathbf{f}_{k_s} q_{k_s},
\end{equation}
where $s_{k_c}\sim\mathcal{CN}(0,1)$ is the information symbol for the $k_c$th communication user and $q_{k_s}\sim\mathcal{CN}(0,1)$ is the probing symbol associated with the $k_s$th sensing target.

In the main framework, the communication--sensing power split is treated as a preset system parameter, while its impact is examined separately in the simulation section. This allows us to focus on the coupled design of antenna deployment, segment partitioning, and beamforming.

\subsection{Communication and Sensing Metrics}\label{subsec:metrics}

For communication, the received signal at communication user $k_c$ is given by
\begin{equation}
	r_{k_c}
	=
	\mathbf{h}_{k_c}^T \mathbf{x} + n_{k_c},
\end{equation}
where $n_{k_c}\sim\mathcal{CN}(0,\sigma_c^2)$ is the additive white Gaussian noise. Therefore, the signal-to-interference-plus-noise ratio (SINR) of user $k_c$ is
\begin{equation}
	\mathrm{SINR}_{k_c}
	=
	\frac{\rho_c |\mathbf{h}_{k_c}^T \mathbf{w}_{k_c}|^2}
	{\rho_c \sum_{i\neq k_c} |\mathbf{h}_{k_c}^T \mathbf{w}_{i}|^2
		+\rho_s \sum_{j=1}^{K_s} |\mathbf{h}_{k_c}^T \mathbf{f}_{j}|^2
		+\sigma_c^2}.
\end{equation}
Accordingly, the achievable sum rate is expressed as
\begin{equation}
	R_{\mathrm{sum}}
	=
	\sum_{k_c=1}^{K_c} \log_2\big(1+\mathrm{SINR}_{k_c}\big).
\end{equation}

For sensing, we adopt the position estimation accuracy of the sensing targets as the performance metric. Specifically, let $\mathbf{J}_{k_s}$ denote the Fisher information matrix (FIM) associated with the estimation of the $k_s$th target position. Then the corresponding Cram\'er--Rao lower bound (CRLB) matrix is given by $\mathbf{J}_{k_s}^{-1}$, and the sensing accuracy metric is defined as
\begin{equation}
	\mathrm{CRLB}_{k_s}
	\triangleq
	\mathrm{tr}\!\left(\mathbf{J}_{k_s}^{-1}\right),
\end{equation}
where a smaller value indicates better sensing accuracy. The detailed derivation of the adopted position CRLB is provided in Appendix~\ref{app:crlb}.

The communication and sensing objectives are inherently coupled. Increasing communication power or steering beams more aggressively toward users may improve $R_{\mathrm{sum}}$, but it may also reduce the effective sensing aperture or degrade the estimation accuracy of the sensing targets. This tradeoff is further affected by the antenna deployment and segment-wise Tx/Rx partition.

\begin{figure*}[t!]
	\centering
	\captionsetup{justification=raggedright,singlelinecheck=false}
	\includegraphics[width=0.6\textwidth]{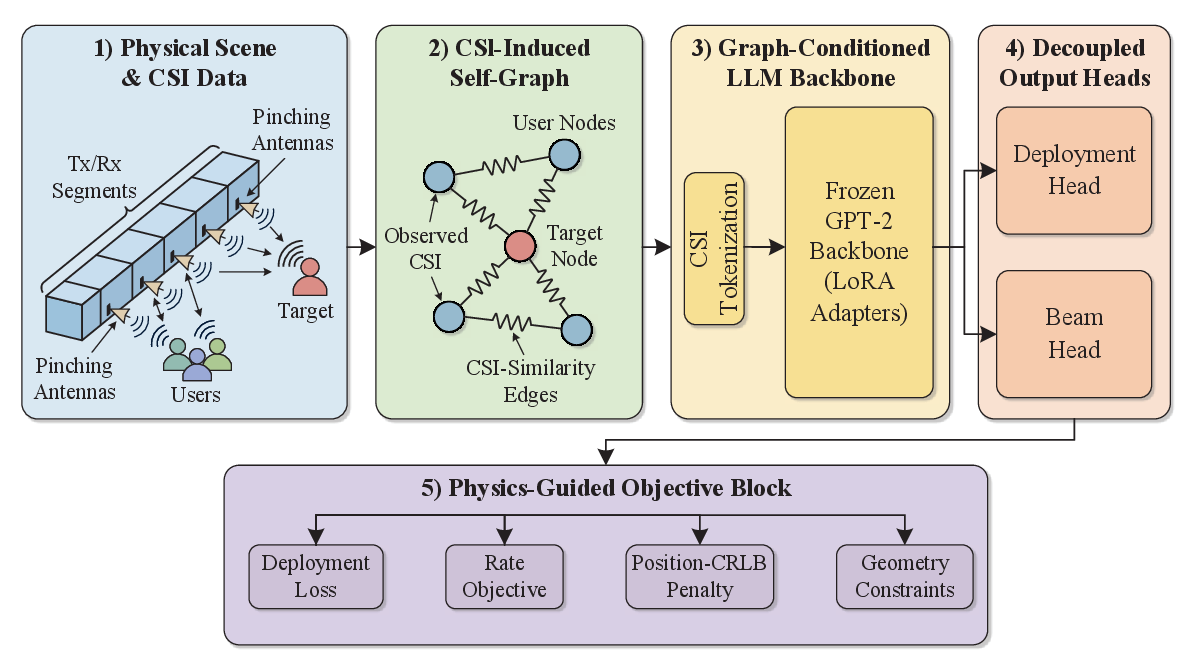}
	\caption{The proposed LLM-enabled general learning framework for SWAN-ISAC.}
	\label{Fig.2}
\end{figure*}

\subsection{Problem Formulation}\label{subsec:opt}

Based on the above model, we aim to jointly optimize the antenna deployment vector $\mathbf{y}$, the segment-wise Tx/Rx partition vector $\boldsymbol{\chi}=[\chi_1,\ldots,\chi_M]^T$, and the communication/sensing beamforming variables $\{\mathbf{w}_{k}\}_{k\in\mathcal{K}_c}$ and $\{\mathbf{f}_{\ell}\}_{\ell\in\mathcal{K}_s}$ so as to maximize the communication utility while satisfying sensing and physical-feasibility requirements. Here, $\mathcal{K}_c$ and $\mathcal{K}_s$ denote the sets of active communication users and sensing targets in the considered system realization, and their cardinalities may vary across different realizations. The resulting problem can be formulated as
\begin{subequations}\label{eq:opt_problem}
	\begin{align}
		\max_{\mathbf{y},\,\boldsymbol{\chi},\,\{\mathbf{w}_{k}\},\,\{\mathbf{f}_{\ell}\}}
		&\quad R_{\mathrm{sum}} \notag \\
		\text{s.t.}\quad
		&\quad \mathrm{CRLB}_{\ell} \le \varepsilon_{\ell}, \quad \forall \ell\in\mathcal{K}_s, \\
		&\quad 0 \le y_n \le L, \quad \forall n, \\
		&\quad |y_n-y_{n'}| \ge d_{\min}, \quad \forall n\neq n', \\
		&\quad \chi_m \in \{0,1\}, \quad \forall m, \\
		&\quad |\mathcal{K}_c| + |\mathcal{K}_s| \le \sum_{m=1}^{M}\chi_m \le M-1, \\
		&\quad \rho_c \sum_{k\in\mathcal{K}_c}\|\mathbf{w}_{k}\|_2^2
		+\rho_s \sum_{\ell\in\mathcal{K}_s}\|\mathbf{f}_{\ell}\|_2^2
		\le P_{\max}.
	\end{align}
\end{subequations}
where $\varepsilon_{\ell}$ is the maximum tolerable sensing error threshold for target $\ell$, and $P_{\max}$ is the total transmit power budget.

Problem \eqref{eq:opt_problem} is highly challenging due to the strong coupling among deployment, partitioning, and beamforming. In particular, the segment-wise Tx/Rx partition introduces discrete structural decisions, whereas the deployment and beamforming variables are continuous and jointly affect both communication rate and sensing CRLB. Moreover, since the active user set $\mathcal{K}_c$ and target set $\mathcal{K}_s$ may vary across different system realizations, directly solving \eqref{eq:opt_problem} for each realization becomes computationally prohibitive. These observations motivate the learning-based structured design developed in the next section, where the mapping from channel state information to deployment, partitioning, and beamforming decisions is learned in a unified framework.

\section{Proposed General Learning Framework}

This section presents the proposed general learning framework for SWAN-ISAC systems. As discussed in Section~\ref{sec:system}, the considered design problem is highly coupled, since the antenna deployment, segment-wise transmit/receive partition, and communication/sensing beamforming jointly affect both the communication sum rate and the sensing CRLB. Moreover, the numbers of communication users and sensing targets may vary across scenarios, which makes conventional fixed-input learning architectures difficult to generalize. To address these challenges, we develop a graph-enhanced LLM-based framework that directly learns the mapping from channel state information (CSI) to deployment, partitioning, and beamforming decisions.
The proposed framework consists of three main components:
 
 1) a CSI-induced self-graph encoder for permutation-invariant scenario representation; 
 
 2) an LLM backbone with LoRA adaptation for high-capacity sequence modeling;
 
 3) two task-specific output heads for deployment/partition prediction and beamforming prediction, respectively. 
 
 In addition, a lightweight transfer mechanism is introduced to support changed user configurations by reusing the learned deployment representation while adapting only the beamforming head. Algorithm~\ref{alg:training_framework_short} summarizes the training procedure of the proposed framework, while Algorithm~\ref{alg:beam_head_transfer_short} summarizes the beam-head-only transfer strategy under changed user counts.

\subsection{Framework Overview}

Let the CSI tensor corresponding to one scenario be denoted by
\begin{equation}
	\mathbf{H}\in\mathbb{R}^{(K_c+K_s)\times N\times 2},
\end{equation}
where the last dimension corresponds to the real and imaginary parts of the complex channel coefficients. Given $\mathbf{H}$, the objective of the proposed framework is to jointly predict
\begin{equation}
	\{\hat{\mathbf{y}}, \hat{\boldsymbol{\chi}}, \hat{\mathbf{W}}, \hat{\mathbf{F}}\},
\end{equation}
where $\hat{\mathbf{y}}$ is the predicted antenna deployment vector, $\hat{\boldsymbol{\chi}}$ is the segment-wise Tx/Rx partition, and $\hat{\mathbf{W}}$ and $\hat{\mathbf{F}}$ denote the communication and sensing beamforming variables, respectively.
The proposed framework first transforms the CSI into a self-graph representation that captures the interaction structure among communication users and sensing targets. The graph-enhanced representation is then processed by an LLM backbone, followed by two task-specific prediction heads. The deployment head is mainly responsible for learning the hierarchical deployment of SWAN, while the beamforming head predicts the beamforming vectors for users and targets. Finally, the LoRA adapters and decoupled output heads are updated according to four physics-guided objectives: deployment loss, rate, CRLB, and geometry constraints.

\subsection{CSI-Induced Self-Graph Encoder}\label{subsec:selfgraph}

\subsubsection{Graph Construction}

To capture the relational structure of the current scenario, we construct a graph
\begin{equation}
	\mathcal{G}=(\mathcal{V},\mathcal{E},\mathbf{A}),
\end{equation}
whose node set is defined as
\begin{equation}
	\mathcal{V}=\{v_1,\ldots,v_{K_c+K_s}\},
\end{equation}
where the first $K_c$ nodes correspond to communication users and the remaining $K_s$ nodes correspond to sensing targets.

For node $i$, let $\mathbf{h}_i\in\mathbb{C}^{N\times 1}$ denote its complex channel vector. The node feature is constructed as
\begin{equation}
	\mathbf{x}_i
	=
	\left[
	\|\mathbf{h}_i\|_2,\;
	\angle\!\left(\sum_{n=1}^{N} [\mathbf{h}_i]_n\right),\;
	\tau_i
	\right]^T,
\end{equation}
where $\tau_i\in\{0,1\}$ is a role indicator distinguishing communication and sensing nodes.

Different from conventional GNNs defined on a fixed graph, the adjacency matrix in the proposed framework is generated adaptively from the CSI of the current sample. Specifically, the edge weight between nodes $i$ and $j$ is defined as
\begin{equation}
	A_{ij}
	=
	\frac{\left|\mathbf{h}_i^H \mathbf{h}_j\right|}
	{\|\mathbf{h}_i\|_2 \|\mathbf{h}_j\|_2 + \epsilon},
\end{equation}
where $\epsilon>0$ is a small constant for numerical stability. After row normalization, the resulting adjacency matrix reflects the CSI-induced interaction pattern among communication users and sensing targets in the current scenario.

This construction is particularly suitable for the considered problem. First, it avoids relying on a hand-crafted fixed graph topology. Second, since the graph is built directly from CSI similarity, it naturally adapts to varying user locations and target realizations. Third, by treating users and targets as graph nodes, the resulting representation is permutation-invariant and scalable to different node cardinalities.

\subsubsection{Graph Propagation and Scenario Embedding}

Given the node features and adjacency matrix, a multi-layer self-graph encoder is applied to aggregate scenario-dependent relational information. Let $\mathbf{h}_i^{(0)}=\mathbf{W}_{\mathrm{in}}\mathbf{x}_i$ denote the projected input feature of node $i$. Then, for graph layer $\ell$, the hidden representation is updated as
\begin{equation}
	\mathbf{h}_i^{(\ell+1)}
	=
	\sigma\!\left(
	\mathbf{W}_{1}^{(\ell)} \mathbf{h}_i^{(\ell)}
	+
	\mathbf{W}_{2}^{(\ell)}
	\sum_{j=1}^{K_c+K_s} A_{ij}\mathbf{h}_j^{(\ell)}
	\right),
\end{equation}
where $\sigma(\cdot)$ denotes the nonlinear activation function.

After $L_g$ graph propagation layers, the global scenario embedding is obtained by mean pooling:
\begin{equation}
	\mathbf{z}_g
	=
	\frac{1}{K_c+K_s}
	\sum_{i=1}^{K_c+K_s}\mathbf{h}_i^{(L_g)}.
\end{equation}
The resulting embedding captures the interaction structure of the current communication-sensing scenario and serves as a permutation-invariant global descriptor for the downstream LLM backbone.

\begin{algorithm}[t]
	\caption{Training Procedure of the Proposed Framework}
	\label{alg:training_framework_short}
	\begin{algorithmic}[1]
		\REQUIRE Training dataset $\mathcal{D}$, system configuration $\mathcal{C}$.
		\ENSURE Trained model parameters.
		
		\STATE Initialize the CSI-induced self-graph encoder, the shared LLM backbone with LoRA adaptation, the deployment head, and the beamforming head.
		\FOR{each training epoch}
		\FOR{each mini-batch $(\mathbf{H}, \mathbf{y}^{\star}, \{\mathbf{p}_{k_c}\}, \{\mathbf{p}_{k_s}\}) \in \mathcal{D}$}
		\STATE Construct the CSI-induced self-graph and obtain the graph embedding $\mathbf{z}_g$.
		\STATE Tokenize $\mathbf{H}$, inject $\mathbf{z}_g$, and obtain the shared hidden representation $\mathbf{H}_{\mathrm{LLM}}$.
		\STATE Predict $\hat{\mathbf{y}}$ and $\hat{\boldsymbol{\chi}}$ using the deployment head.
		\STATE Predict $\hat{\mathbf{W}}$ and $\hat{\mathbf{F}}$ using the beamforming head.
		\STATE Evaluate $R_{\mathrm{sum}}$ and $\mathrm{CRLB}$ in the differentiable SWAN-ISAC environment.
		\STATE Compute the total loss
		
		$\mathcal{L}=w_{\mathrm{dep}}\mathcal{L}_{\mathrm{dep}}+\mathcal{L}_{\mathrm{perf}}
		+\mathcal{L}_{\mathrm{geom}}$.
		\STATE Update the trainable parameters by backpropagation.
		\ENDFOR
		\ENDFOR
	\end{algorithmic}
\end{algorithm}

\subsection{LLM Backbone With LoRA Fine-tuning}\label{subsec:llm}

Although the graph embedding $\mathbf{z}_g$ already provides a compact scenario summary, the considered SWAN-ISAC design still involves highly coupled structural and beamforming decisions. To model these couplings more effectively, we employ an LLM-style backbone as a high-capacity sequence encoder.

First, the numerical CSI tensor is converted into a token sequence. Specifically, for the $n$th antenna position, the real and imaginary CSI coefficients associated with all users and targets are concatenated into one token feature vector. Denoting the resulting token sequence by
\begin{equation}
	\mathbf{X}=[\mathbf{x}^{\mathrm{tok}}_1,\ldots,\mathbf{x}^{\mathrm{tok}}_N]^T,
\end{equation}
each token is projected into a hidden space as
\begin{equation}
	\tilde{\mathbf{x}}_n
	=
	\mathrm{LN}\!\left(\mathbf{W}_{\mathrm{tok}}\mathbf{x}^{\mathrm{tok}}_n\right),
\end{equation}
where $\mathrm{LN}(\cdot)$ denotes layer normalization.

The graph embedding is then injected into the token sequence through an additive conditioning mechanism:
\begin{equation}
	\bar{\mathbf{x}}_n
	=
	\tilde{\mathbf{x}}_n
	+
	\mathbf{W}_{g}\mathbf{z}_g,
	\quad n=1,\ldots,N.
\end{equation}
In this way, each CSI token is enhanced by the global relational information extracted from the self-graph.

The resulting token sequence is fed into an LLM backbone:
\begin{equation}
	\mathbf{H}_{\mathrm{LLM}}
	=
	\mathrm{LLM}_{\mathrm{LoRA}}(\bar{\mathbf{X}}),
\end{equation}
where $\mathrm{LLM}_{\mathrm{LoRA}}(\cdot)$ denotes the LLM encoder with LoRA adaptation. In the proposed framework, LoRA modules are inserted into the main projection layers of the LLM backbone so that most backbone parameters can remain frozen while only lightweight low-rank adapters are updated. This design preserves the expressive capacity of the backbone while reducing the number of trainable parameters and improving adaptation efficiency.

Compared with directly applying a standard multilayer perceptron or a conventional transformer encoder, the proposed graph-enhanced LLM backbone provides two advantages. First, it benefits from the permutation-invariant and scenario-aware self-graph representation. Second, it offers a stronger capacity for modeling the nonlinear coupling among deployment, partitioning, and beamforming decisions.

\subsection{Two Task-Specific Prediction Heads}\label{subsec:heads}

To align the model structure with the considered SWAN-ISAC design problem, the shared hidden representation produced by the graph-enhanced LLM backbone is decoded by two task-specific heads. One head focuses on structural prediction, namely deployment and segment-wise partitioning, while the other focuses on communication/sensing beamforming prediction.

\subsubsection{Deployment and Partition Head}

Let $\mathbf{H}_{\mathrm{LLM}}\in\mathbb{R}^{N\times d}$ denote the hidden token sequence from the shared backbone, where $d$ is the hidden dimension. A pooled representation for structural prediction is first obtained as
\begin{equation}
	\mathbf{g}_{\mathrm{dep}}
	=
	f_{\mathrm{dep}}\!\left(
	\frac{1}{N}\sum_{n=1}^{N}\mathbf{H}_{\mathrm{LLM}}(n,:)
	\right),
\end{equation}
where $f_{\mathrm{dep}}(\cdot)$ denotes a lightweight projection module.

The deployment head outputs
\begin{equation}
	[\hat{\mathbf{y}}^{\mathrm{raw}}, \bm{\pi}_{\chi}]
	=
	g_{\mathrm{dep}}(\mathbf{g}_{\mathrm{dep}}),
\end{equation}
where $\hat{\mathbf{y}}^{\mathrm{raw}}\in\mathbb{R}^{N}$ denotes the raw deployment prediction and $\bm{\pi}_{\chi}\in\mathbb{R}^{M}$ denotes the segment-wise partition logits.

The final deployment prediction is obtained through
\begin{equation}
	\hat{\mathbf{y}}
	=
	L \cdot \sigma(\hat{\mathbf{y}}^{\mathrm{raw}}),
\end{equation}
where $\sigma(\cdot)$ is the sigmoid function. To ensure physical feasibility, the predicted deployment is further processed by a continuous non-overlap projection operator, which enforces range and spacing consistency while maintaining differentiability during training.

The segment-wise transceiver partition is obtained by a top-$K$ operation on the logits $\bm{\pi}_{\chi}$. Specifically, the predicted transmit set is formed by selecting the $K_{\mathrm{tx}}$ largest entries of $\bm{\pi}_{\chi}$, where
\begin{equation}
	K_{\mathrm{tx}} = K_c + K_s
\end{equation}
in the default setting. The resulting binary partition vector is denoted by $\hat{\boldsymbol{\chi}}$.

This design is more compact than directly predicting a full element-wise binary activation mask. It also better matches the structural characteristics of the SWAN system, where the key design decisions are the antenna deployment and the segment-wise Tx/Rx partition.

\subsubsection{Beamforming Head}

In parallel with the deployment head, the beamforming head uses another pooled hidden representation
\begin{equation}
	\mathbf{g}_{\mathrm{bf}}
	=
	f_{\mathrm{bf}}\!\left(
	\frac{1}{N}\sum_{n=1}^{N}\mathbf{H}_{\mathrm{LLM}}(n,:)
	\right)
\end{equation}
to predict the communication and sensing beamforming variables. Specifically, the beamforming head outputs
\begin{equation}
	[\hat{\mathbf{W}}_{\Re}, \hat{\mathbf{W}}_{\Im},
	\hat{\mathbf{F}}_{\Re}, \hat{\mathbf{F}}_{\Im}]
	=
	g_{\mathrm{bf}}(\mathbf{g}_{\mathrm{bf}}),
\end{equation}
where the real and imaginary parts are combined to form the complex beamforming matrices
\begin{align}
	\hat{\mathbf{W}}
	&=
	\hat{\mathbf{W}}_{\Re}
	+
	j\hat{\mathbf{W}}_{\Im},\\
	\hat{\mathbf{F}}
	&=
	\hat{\mathbf{F}}_{\Re}
	+
	j\hat{\mathbf{F}}_{\Im}.
\end{align}

The predicted beamforming variables are then evaluated in the differentiable SWAN-ISAC environment together with the deployment prediction and the segment-wise partition. Therefore, the beamforming head is trained jointly with the deployment head under the same communication-sensing objective.

\begin{algorithm}[t]
	\caption{Beam-Head-Only Transfer Under Changed User Count}
	\label{alg:beam_head_transfer_short}
	\begin{algorithmic}[1]
		\REQUIRE Trained source model $\mathcal{M}_{\mathrm{src}}$, target configuration $(K_c^{\mathrm{tgt}}, K_s^{\mathrm{tgt}})$, target dataset $\mathcal{D}_{\mathrm{tgt}}$.
		\ENSURE Adapted target model $\mathcal{M}_{\mathrm{tgt}}$.
		
		\STATE Load $\mathcal{M}_{\mathrm{src}}$ and reuse the self-graph encoder, the shared LLM backbone, and the deployment head.
		\STATE Obtain the transferred deployment prediction $\hat{\mathbf{y}}_{\mathrm{transfer}}$ by deployment-only inference.
		\STATE Reset the beamforming head according to $(K_c^{\mathrm{tgt}}, K_s^{\mathrm{tgt}})$.
		\STATE Freeze the self-graph encoder, the shared LLM backbone, and the deployment head.
		\FOR{each training epoch on $\mathcal{D}_{\mathrm{tgt}}$}
		\FOR{each mini-batch $(\mathbf{H}, \{\mathbf{p}_{k_c}\}, \{\mathbf{p}_{k_s}\}) \in \mathcal{D}_{\mathrm{tgt}}$}
		\STATE Reuse $\hat{\mathbf{y}}_{\mathrm{transfer}}$ and predict $\hat{\mathbf{W}}, \hat{\mathbf{F}}$ using the beamforming head.
		\STATE Evaluate $R_{\mathrm{sum}}$ and $\mathrm{CRLB}$ and update only the beamforming head.
		\ENDFOR
		\ENDFOR
	\end{algorithmic}
\end{algorithm}

\subsection{User-Count Transfer Mechanism}\label{subsec:transfer}

A key objective of this work is to examine whether the learned deployment policy is site-specific and thus reusable when the communication user count changes. To this end, the proposed framework is designed with an explicit transfer mechanism.

The main intuition is that the deployment branch should primarily capture the structural characteristics of the site and the segmented aperture, whereas the beamforming branch should absorb most of the user-dependent adaptation burden. Accordingly, when transferring a source model trained under one user configuration to a target scenario with a different number of communication users, the encoder, shared LLM backbone, and deployment head are retained, while only the beamforming head is reset and adapted.

More specifically, the transfer procedure includes the following steps:
\begin{itemize}
	\item STEP 1: Train the source model on a source user configuration and obtain the learned encoder, LLM backbone, deployment head, and beamforming head.
	\item STEP 2: For a target task with changed user count, reuse the encoder, backbone, and deployment head to generate the transferred deployment prediction.
	\item STEP 3: Reset the beamforming output head according to the new output dimension and fine-tune only this head on the target task.
\end{itemize}

If the transferred deployment remains stable while the target-task communication and sensing performance can be recovered by adapting only the beamforming head, then this provides evidence that the learned deployment policy is tied to the site structure rather than overfitting one fixed user configuration.

\subsection{Training Objective}\label{subsec:training_obj}

The proposed framework is trained by combining deployment supervision, communication-sensing utility optimization, and physical-feasibility regularization. The overall loss is written as
\begin{equation}
	\mathcal{L}
	=
	w_{\mathrm{dep}}\mathcal{L}_{\mathrm{dep}}
	+
	\mathcal{L}_{\mathrm{perf}}
	+
	\mathcal{L}_{\mathrm{geom}}
,
\end{equation}
where each term is defined as follows.

\subsubsection{Deployment Loss}

To supervise the structural prediction, the deployment loss is defined as
\begin{equation}
	\mathcal{L}_{\mathrm{dep}}
	=
	\frac{1}{N}
	\left\|
	\frac{\mathrm{sort}(\hat{\mathbf{y}})-\mathrm{sort}(\mathbf{y}^{\star})}{L}
	\right\|_2^2,
\end{equation}
where $\mathbf{y}^{\star}$ denotes the reference deployment label. The sorting operation makes the loss permutation-invariant with respect to antenna indexing.

\subsubsection{Communication-Sensing Performance Loss}

The communication-sensing performance loss is designed to maximize the communication sum rate while penalizing sensing-quality violation:
\begin{equation}
	\small
	\mathcal{L}_{\mathrm{perf}}
	=
	-
	w_{\mathrm{rate}} R_{\mathrm{sum}}
	+
	w_{\mathrm{crlb}}
	\max\!\left(
	0,\;
	\log(\mathrm{CRLB}+\epsilon)-\log(\varepsilon_{\mathrm{crlb}})
	\right),
\end{equation}
where $\varepsilon_{\mathrm{crlb}}$ denotes the sensing threshold and $\epsilon$ is a small constant used for numerical stability. The logarithmic hinge form improves optimization stability when the CRLB spans several orders of magnitude.

\subsubsection{Geometry Regularization}

To ensure physically meaningful deployment predictions, a geometry regularization term $\mathcal{L}_{\mathrm{geom}}$ is imposed, which considers spacing violations, encourages valid segment coverage, and penalizes out-of-range deployment predictions.

\subsection{Training and Complexity Discussion}\label{subsec:train_complexity}

All trainable components are optimized jointly using gradient-based learning. During source-task training, both task-specific heads and the trainable backbone components are updated. During transfer, only the beamforming head is adapted, which substantially reduces the number of trainable parameters and convergence cost.

In terms of complexity, the self-graph encoder mainly introduces pairwise interaction modeling among communication users and sensing targets, which scales as
\begin{equation}
	\mathcal{O}\big((K_c+K_s)^2\big).
\end{equation}
The LLM backbone dominates the sequence modeling cost, while the task-specific heads contribute comparatively low additional complexity. Therefore, compared with lightweight baselines such as multilayer perceptrons, the proposed framework incurs higher per-run computation cost. However, this extra cost is compensated by its stronger structural representation capability and, more importantly, its parameter-efficient transfer behavior under changed user configurations.

\begin{figure*}[t]
	\centering
	
	\begin{subfigure}[b]{0.48\textwidth}
		\centering
		\includegraphics[width=\linewidth]{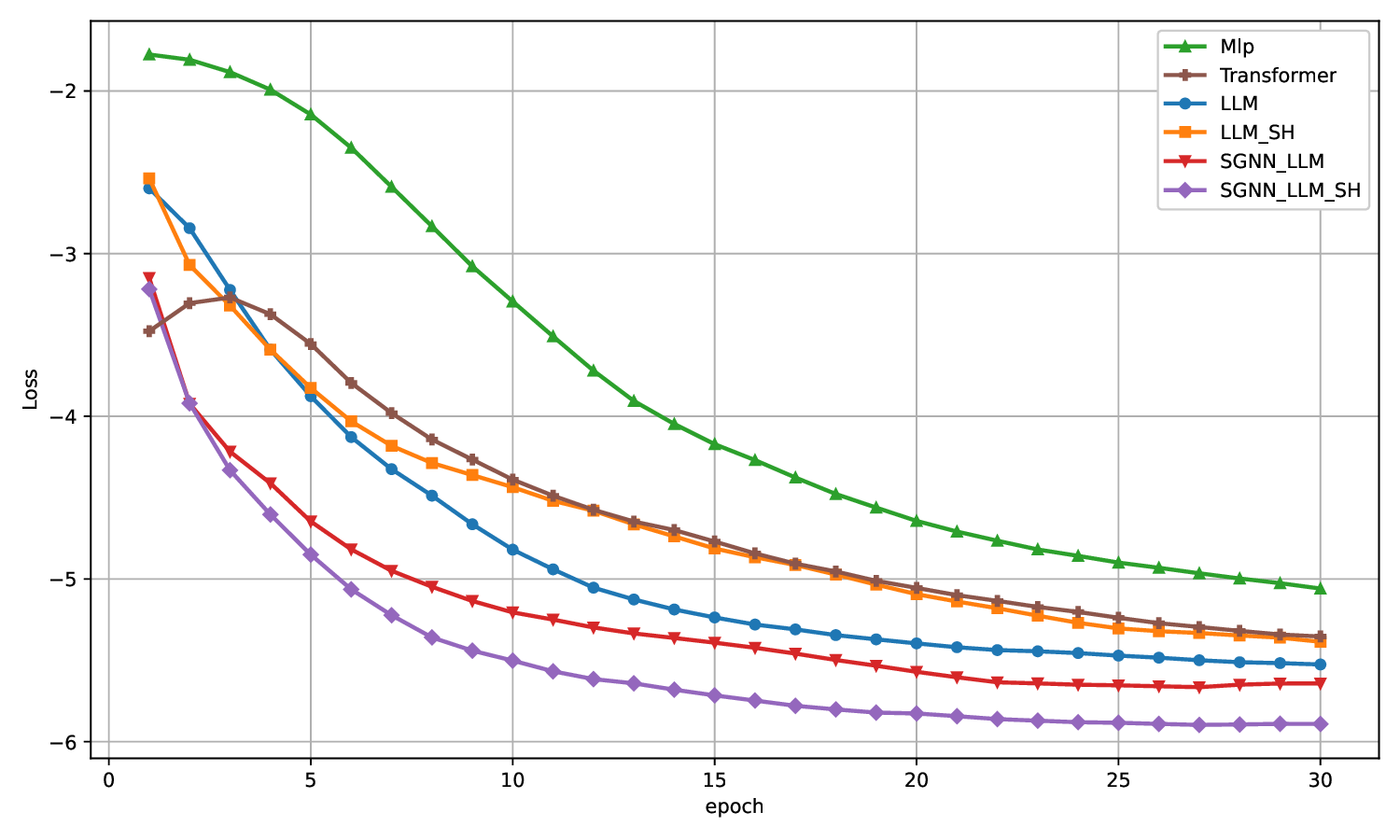}
		\caption{Validation loss versus epoch.}
		\label{fig:main_loss}
	\end{subfigure}
	\hfill
	\begin{subfigure}[b]{0.48\textwidth}
		\centering
		\includegraphics[width=\linewidth]{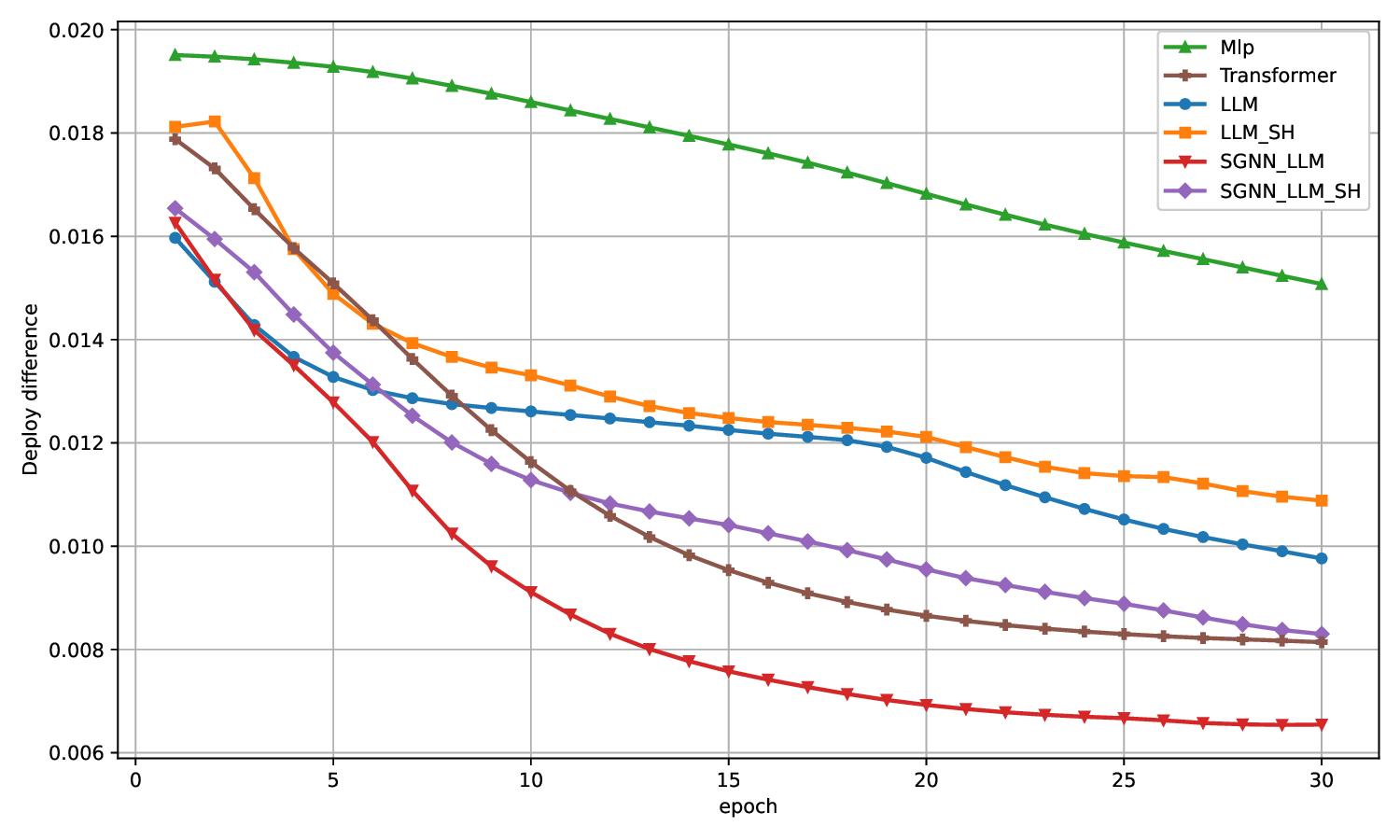}
		\caption{Validation deployment difference versus epoch.}
		\label{fig:main_dep}
	\end{subfigure}
	
	\vspace{0.25cm}
	
	\begin{subfigure}[b]{0.48\textwidth}
		\centering
		\includegraphics[width=\linewidth]{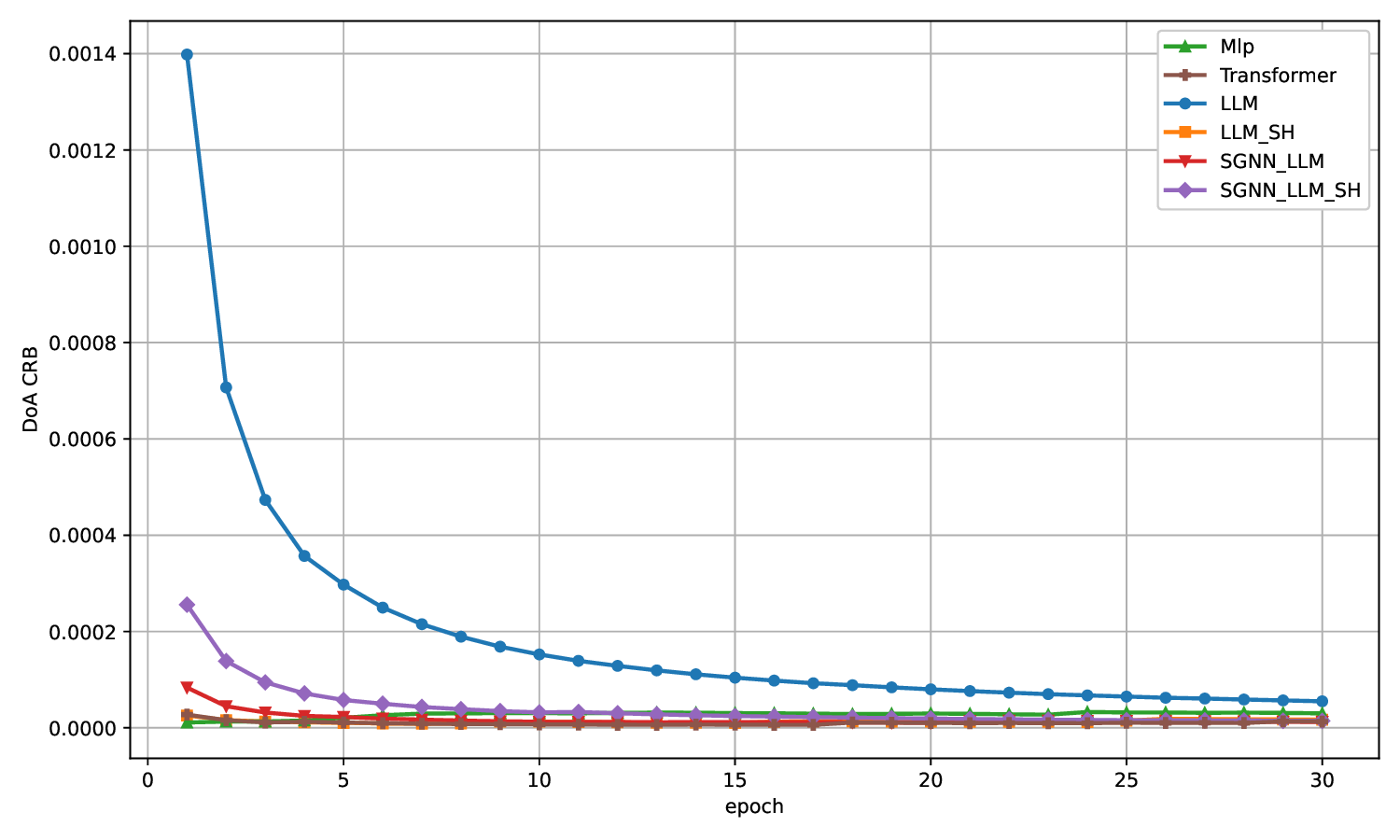}
		\caption{Validation position CRLB versus epoch.}
		\label{fig:main_doa}
	\end{subfigure}
	\hfill
	\begin{subfigure}[b]{0.48\textwidth}
		\centering
		\includegraphics[width=\linewidth]{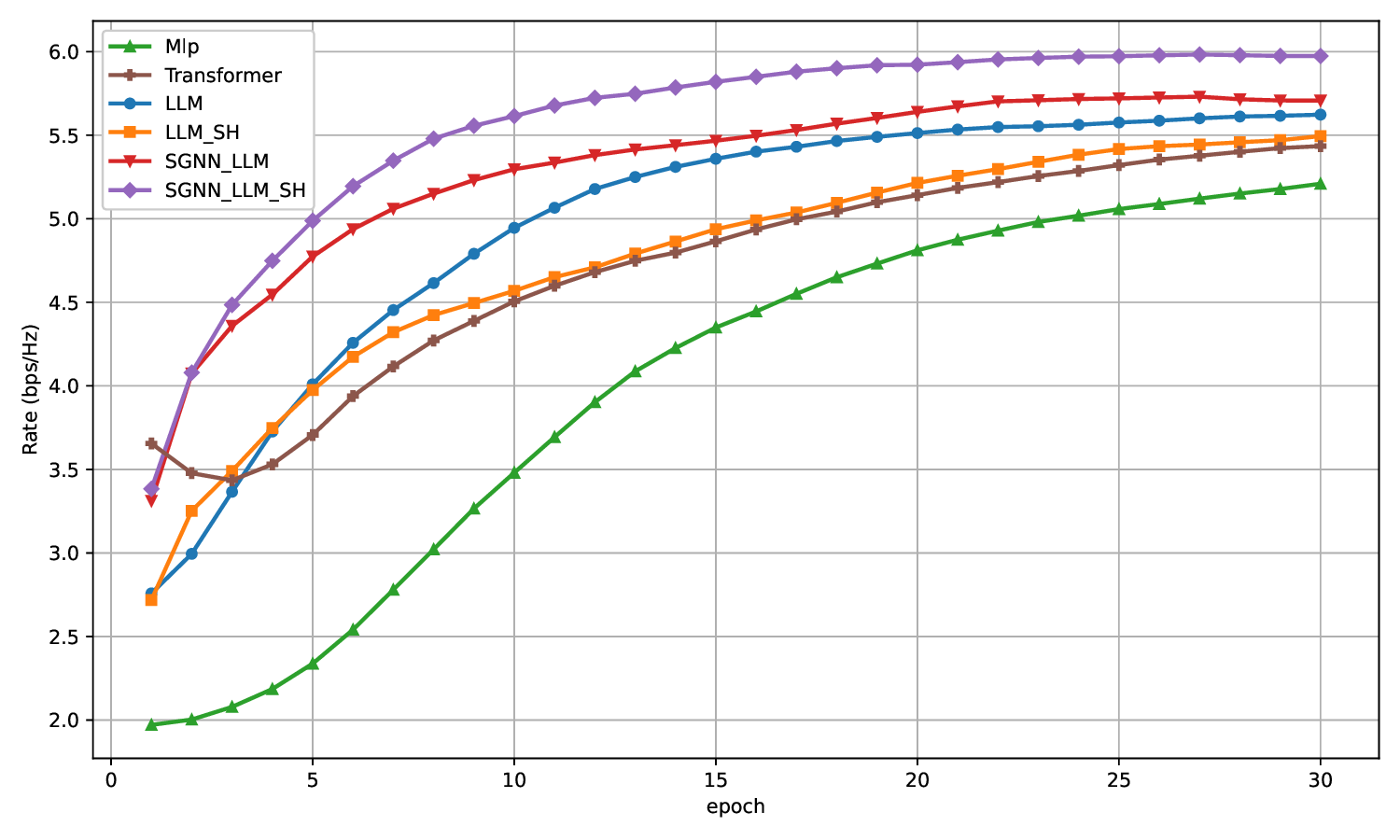}
		\caption{Validation sum rate versus epoch.}
		\label{fig:main_rate}
	\end{subfigure}
	
	\caption{Convergence behavior and main benchmark comparison of different model families.}
	
	\label{Fig.3}
\end{figure*}

\section{Simulation Results}\label{section:4}

In this section, the performance of the proposed framework is evaluated from five perspectives: 1) convergence behavior and benchmark comparison, 2) qualitative visualization of antenna deployment and beam patterns, 3) robustness under different communication--sensing power allocation ratios and imperfect CSI, 4) model complexity, and 5) transferability across different user numbers. Unless otherwise specified, the benchmark experiments are conducted under the settings in table \ref{settings}.

\begin{table}[t]
	\centering

	\begin{tabular}{ll}
		\hline
		\textbf{Parameter} & \textbf{Value} \\
		\hline
		Number of segmented waveguides $M$ & 4 \\
		Number of pinching antennas $N$ & 40 \\
		Number of communication users $K_c$ & 2 \\
		Number of sensing targets $K_s$ & 1 \\
		Waveguide length $L$ & 50 m \\
		Maximum transmit power $P_{\max}$ & 10 \\
		Total channel realizations & 3000 \\
		Training/validation/test split & 70\% / 15\% / 15\% \\
		Learning rate & $5\times 10^{-4}$ \\
		Training batch size & 64 \\
		Evaluation batch size & 128 \\
		Number of training epochs & 30 \\
		Gradient clipping norm threshold & 1.0 \\

		\hline
	\end{tabular}
\caption{Benchmark Experiment Settings}
	\label{settings}
\end{table}

The training objective jointly accounts for communication performance, sensing accuracy, and deployment supervision. In the default setting, the corresponding loss weights are set to $w_{\mathrm{rate}}=1.0$, $w_{\mathrm{CRLB}}=0.2$, and $w_{\mathrm{dep}}=10.0$, respectively. Unless otherwise specified, the communication and sensing power allocation factors are fixed as $\rho_c=0.8$ and $\rho_s=0.2$. For the LLM-based methods, a pretrained GPT-2 backbone is adopted and kept frozen during training, while low-rank adaptation is introduced through LoRA. The LoRA hyperparameters are set to rank $r=32$, scaling factor $\alpha=16$, and dropout rate $0.05$.

The compared methods are summarized as follows. 1) \textit{MLP}: a fully connected baseline that directly flattens the input CSI and regresses the antenna deployment and beamforming variables. 2) \textit{Transformer}: a self-attention-based baseline that treats the antenna-domain CSI as a sequence and performs joint prediction after transformer encoding. 3) \textit{LLM}: an LLM-based baseline that projects the CSI sequence into a pretrained GPT-style backbone with LoRA adaptation for joint deployment and beamforming prediction. 4) \textit{LLM\_SH}: an enhanced LLM variant with decoupled output heads for deployment and beamforming prediction. 5) \textit{SGNN\_LLM}: a hybrid model that incorporates a CSI-induced self-graph encoder before the LLM backbone to capture the structural correlation among communication users and sensing targets. 6) \textit{SGNN\_LLM\_SH}: the proposed method, which combines the self-graph encoder, the LLM backbone, and task-specific decoupled heads to jointly optimize deployment and beamforming decisions.

\subsection{Convergence Behavior and Main Benchmark Comparison}

Fig.~\ref{Fig.3} compares the convergence behavior of different model families under the same training and validation protocol. Several observations can be made. In Fig.~\ref{fig:main_loss}, the proposed SGNN\_LLM\_SH achieves the lowest validation loss among all compared methods and converges steadily throughout the training process. This indicates that combining a CSI-induced self-graph representation with an LLM backbone and two task-specific heads is effective for the coupled SWAN-ISAC design problem. While in Fig.~\ref{fig:main_rate}, SGNN\_LLM\_SH achieves the highest validation sum rate, followed by SGNN\_LLM and LLM-based baselines. This shows that the graph-enhanced LLM design is more effective than purely feedforward or transformer-only baselines in extracting useful structural information from the CSI. In contrast, the MLP baseline improves more slowly and remains consistently inferior in the considered setting, which suggests that shallow ordered-input models are insufficient for the coupled deployment and beamforming design task. Fig.~\ref{fig:main_dep} shows that the deployment metric does not follow exactly the same ranking as the communication rate. In particular, SGNN\_LLM achieves the lowest deployment difference, while SGNN\_LLM\_SH remains competitive but not always the best in this metric. This observation is important because it indicates that the best SWAN-ISAC model should be judged by the overall communication--sensing tradeoff rather than by a single structural metric alone. Fig.~\ref{fig:main_doa} reveals that the plain LLM baseline exhibits a significantly larger sensing CRLB than the other methods during most of the training process, whereas the graph-enhanced methods maintain much more stable sensing behavior. In particular, SGNN\_LLM\_SH avoids the severe CRLB degradation observed in the plain LLM baseline and achieves a favorable balance between high communication rate and reliable sensing accuracy.

Overall, the results in Fig.~\ref{Fig.3} show that SGNN\_LLM\_SH provides the best overall tradeoff among convergence stability, communication performance, and sensing robustness, while SGNN\_LLM is particularly strong in deployment prediction. This confirms that introducing CSI-induced relational modeling before the LLM backbone is beneficial for segmented pinching antenna-assisted ISAC design.

\subsection{Qualitative Visualization of Deployment and Beam Patterns}

\begin{figure}[t!]
	\centering
	\captionsetup{justification=raggedright,singlelinecheck=false}
	\includegraphics[width=0.95\linewidth]{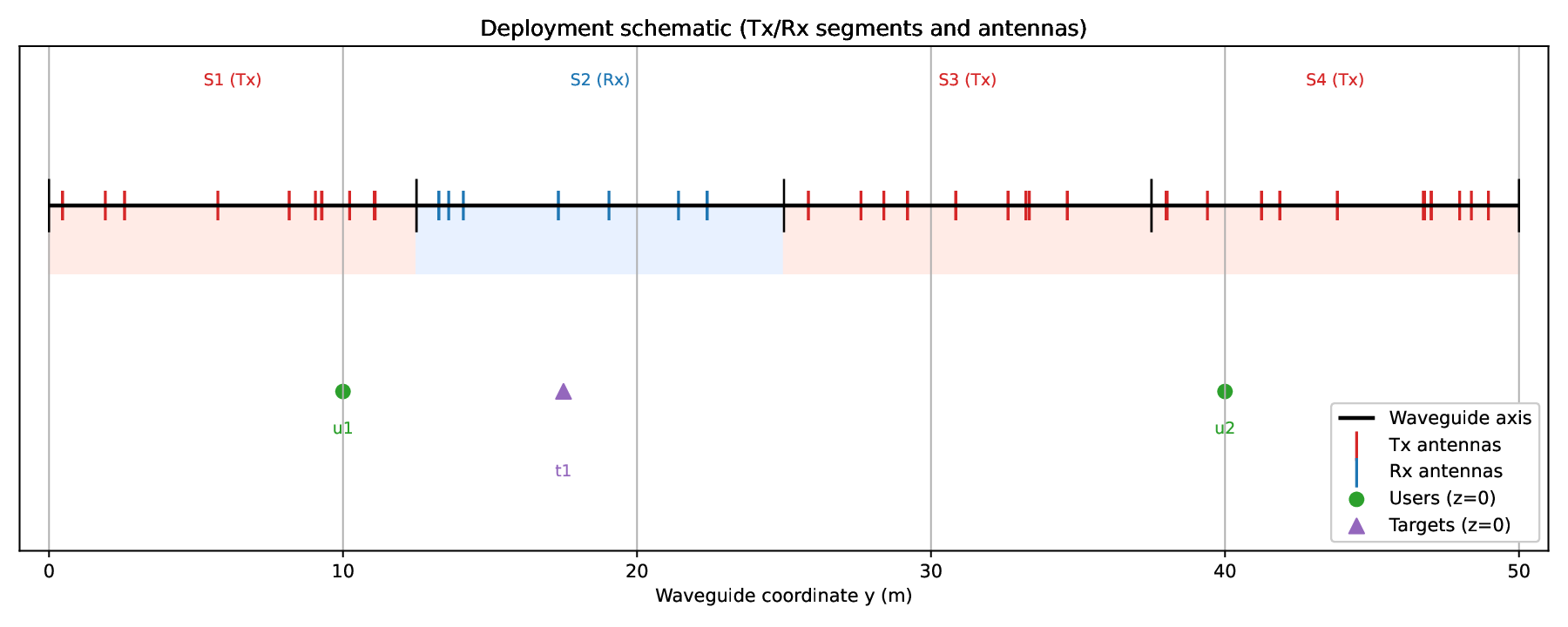}
	\caption{Example of the predicted deployment and segment wise Tx/Rx partition. Red and blue markers indicate antennas assigned to transmit mode and receive mode segments, respectively.}
	\label{Fig.4}
\end{figure}

\begin{figure}[t!]
	\centering
	\captionsetup{justification=raggedright,singlelinecheck=false}
	\includegraphics[width=0.95\linewidth]{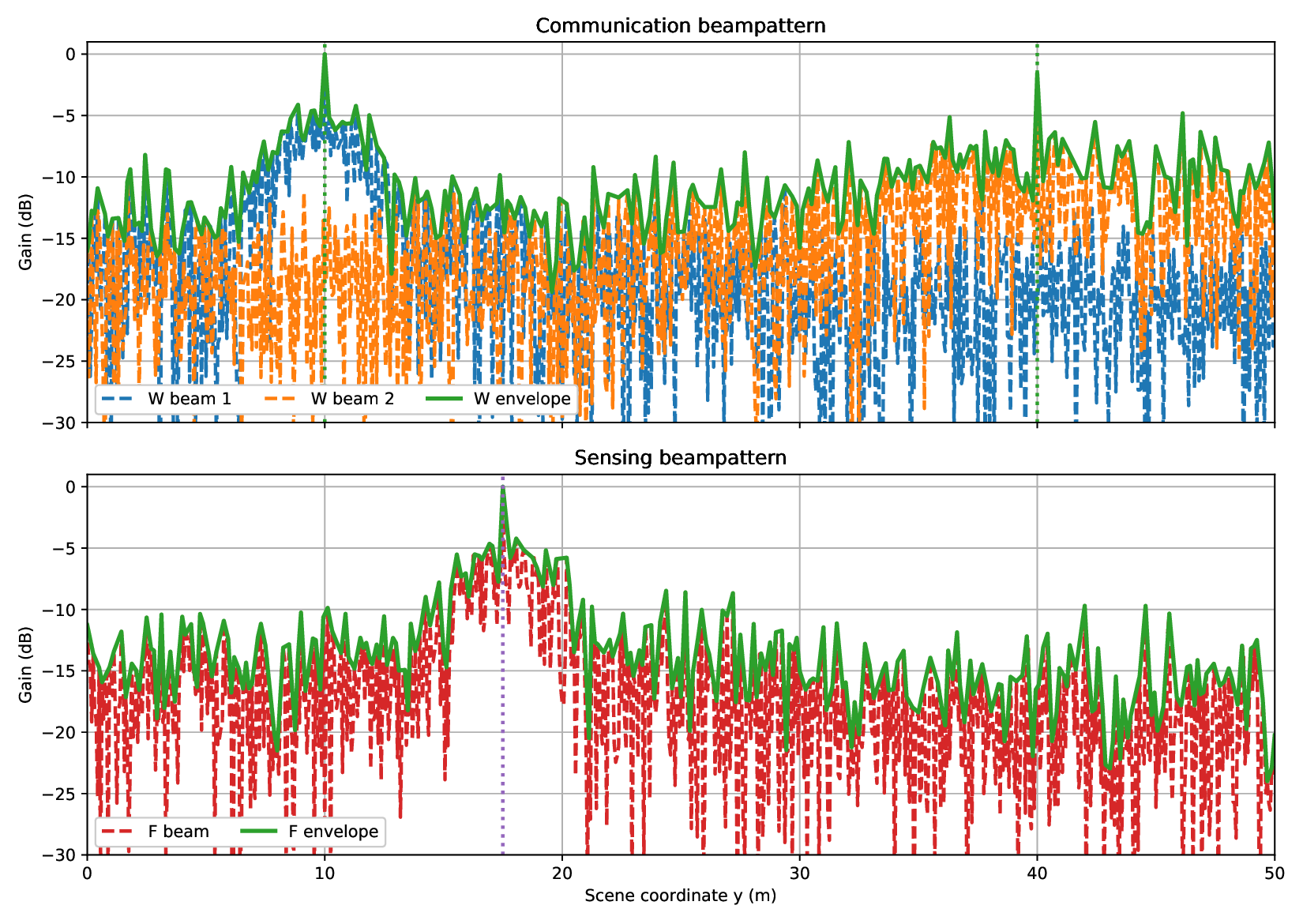}
	\caption{Representative communication and sensing beam patterns produced by the proposed framework. The vertical dotted lines indicate the desired user and target directions in the scene.}
	\label{Fig.5}
\end{figure}

To better understand the learned structure, Fig.~\ref{Fig.4} visualizes one representative deployment predicted by the framework. It can be observed that the learned deployment is compatible with the segmented SWAN structure and produces a clear segment-wise Tx/Rx partition. In the shown example, one segment is configured as a receive segment while the others are used for transmission, which is consistent with the requirement of simultaneously maintaining sensing reception capability and sufficient communication streams. More importantly, the deployed antennas are distributed over the segmented aperture in a physically meaningful manner rather than collapsing into an overly concentrated arrangement.

Fig.~\ref{Fig.5} further illustrates the corresponding communication and sensing beam patterns. In the communication subfigure, the beam envelope exhibits clear enhancement around the desired user locations, while in the sensing subfigure the sensing beam forms a prominent peak around the target location. Although the learned beam patterns are not perfectly idealized, they demonstrate that the proposed framework is able to generate communication and sensing waveforms that are consistent with the predicted structure and the underlying scene geometry. This qualitative result supports the claim that the learned model jointly coordinates deployment, partitioning, and beamforming rather than optimizing them independently.

\subsection{Robustness Under Communication Ratio Variation and Imperfect CSI}

\begin{figure}[t!]
	\centering
	\captionsetup{justification=raggedright,singlelinecheck=false}
	\includegraphics[width=0.9\linewidth]{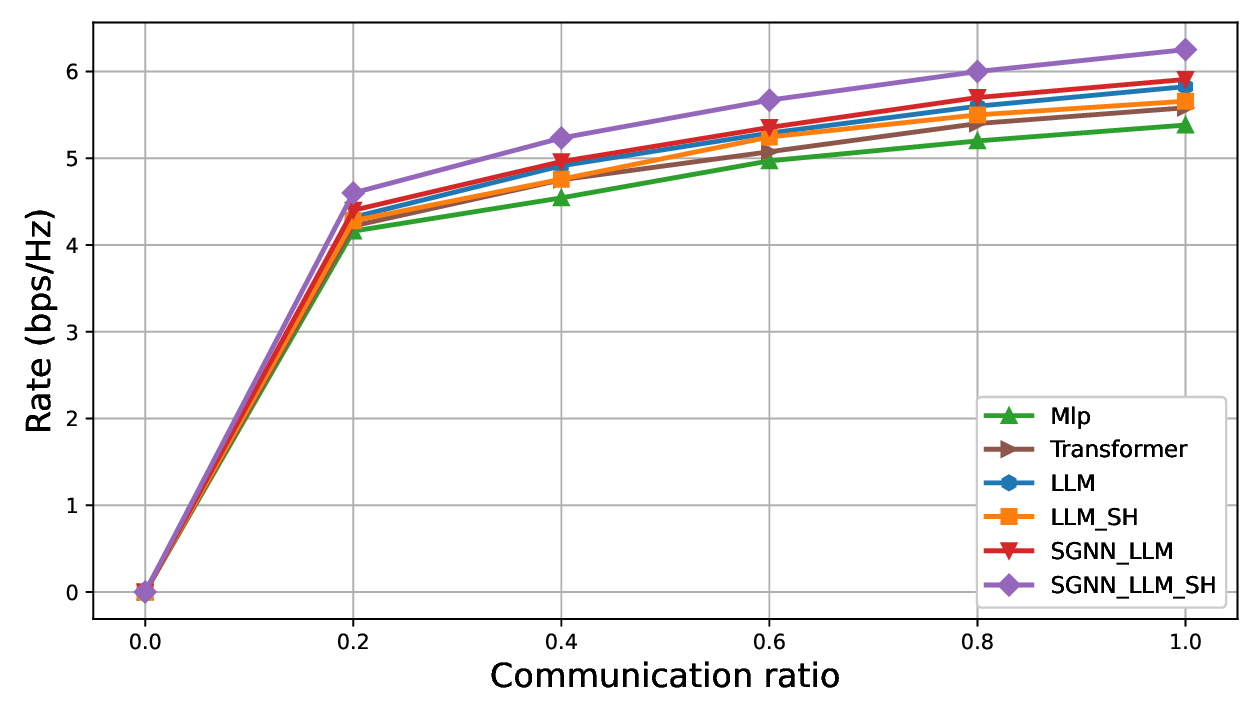}
	\caption{Achievable rate under different communication power ratios.}
	\label{Fig.6}
\end{figure}

\begin{figure}[t!]
	\centering
	\captionsetup{justification=raggedright,singlelinecheck=false}
	\includegraphics[width=0.9\linewidth]{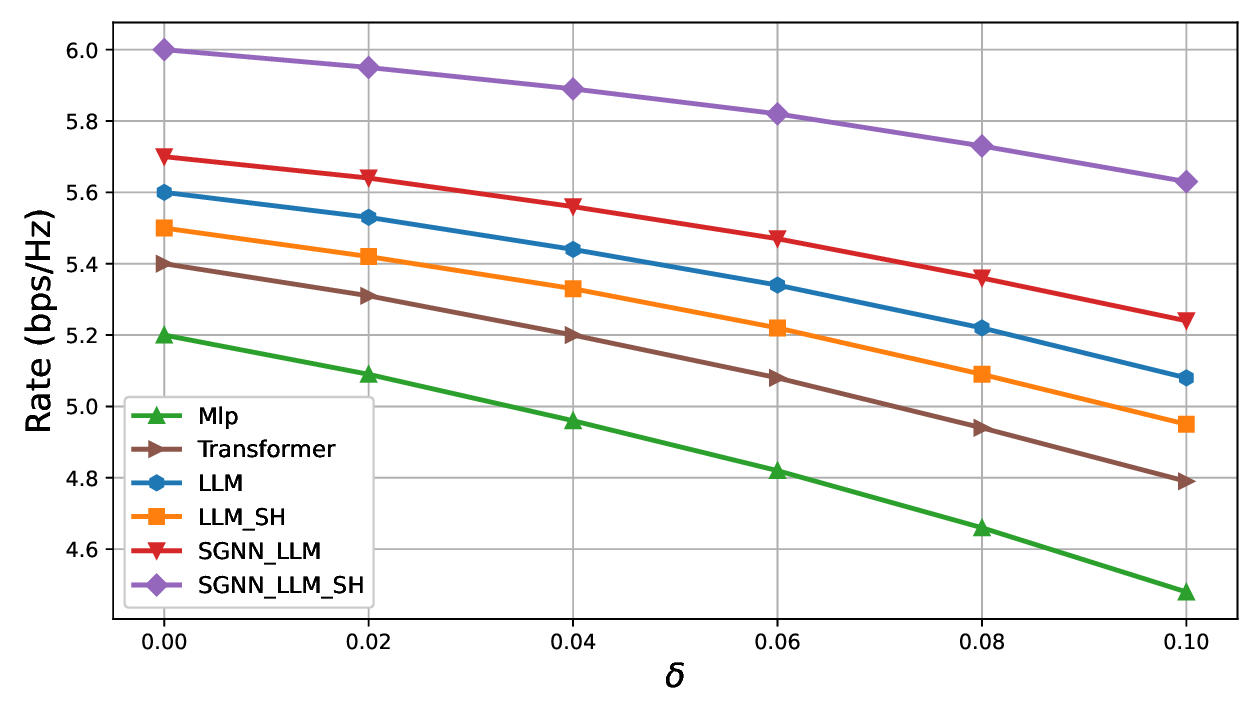}
	\caption{Achievable rate under imperfect CSI with different perturbation levels $\delta$.}
	\label{Fig.7}
\end{figure}

Fig.~\ref{Fig.6} shows the achievable rate under different communication power ratios. Several useful trends can be observed. First, all methods achieve zero rate at $\rho_c=0$, as expected, since no communication power is allocated in that case. Second, as the communication ratio increases, the achievable rate of all methods improves. Third, SGNN\_LLM\_SH consistently achieves the highest rate over the whole range of nonzero communication ratios, indicating that the proposed graph-enhanced LLM structure is robust to different communication--sensing tradeoff settings. In particular, the gain of SGNN\_LLM\_SH becomes more pronounced at moderate-to-high communication ratios, which suggests that the learned representation is effective in converting the available spatial resources into communication performance without severely sacrificing sensing capability.

Fig.~\ref{Fig.7} evaluates robustness under imperfect CSI. As the CSI perturbation level $\delta$ increases, the achievable rate of all methods gradually decreases, which is consistent with the fact that degraded CSI reduces beamforming accuracy. Nevertheless, SGNN\_LLM\_SH remains the strongest method across the whole perturbation range, followed by SGNN\_LLM and the plain LLM baseline. This result indicates that the proposed framework is less sensitive to CSI degradation than the compared baselines. The likely reason is that the CSI-induced self-graph captures relational structure at a higher level than direct ordered-input regression, which makes the learned representation more resilient to moderate perturbations in the raw CSI.

Taken together, Figs.~\ref{Fig.6} and \ref{Fig.7} show that the proposed framework is not limited to one single operating point. Instead, it maintains strong performance under both communication--sensing ratio variation and CSI mismatch, which is important for practical ISAC deployment.

\subsection{Complexity Comparison}

\begin{table*}[t]
	\centering
	\begin{tabular}{lcccccc}
		\toprule
		Metric & MLP & Transformer & LLM & LLM\_SH & SGNN\_LLM & SGNN\_LLM\_SH \\
		\midrule
		Network parameters ($1 \times 10^6$) & 0.07/0.07 & 0.82/0.82 & 5.37/129.81 & 5.96/130.40 & 5.44/129.88 & 6.03/130.47 \\
		Training time (ms) & 3.01 & 33.10 & 364.62 & 343.92 & 361.69 & 347.12 \\
		Inference time (ms) & 0.67 & 4.68 & 123.44 & 125.69 & 120.76 & 123.70 \\
		\bottomrule
	\end{tabular}
	\caption{Comparison of network parameters (training/total), training time, and inference time across different algorithms.}
	\label{tab:algorithm_time_compare}
\end{table*}

Table~\ref{tab:algorithm_time_compare} reports the model size and runtime of different algorithms. As expected, the MLP baseline is by far the lightest model in both parameter count and runtime, while the LLM-based methods are significantly more expensive. The graph-enhanced variants further introduce a moderate increase in complexity due to the additional self-graph encoder and split output design. Nevertheless, the extra complexity of SGNN\_LLM\_SH is relatively small compared with the total LLM backbone scale, while the performance benefits observed in Figs.~\ref{Fig.3}, \ref{Fig.6}, and \ref{Fig.7} are substantial.

Therefore, Table~\ref{tab:algorithm_time_compare} reveals a clear accuracy--efficiency tradeoff. Lightweight baselines are attractive in terms of computation, but they are less competitive in communication rate and robustness. In contrast, SGNN\_LLM\_SH incurs higher computational cost, yet provides a stronger overall communication--sensing tradeoff and, as shown next, offers clear advantages in transferability and adaptation efficiency.

\subsection{User-Count Transfer Verification}

\begin{table*}[t]
	\centering
	\small
	\setlength{\tabcolsep}{7pt}
	\renewcommand{\arraystretch}{1.12}
	\begin{tabular}{llccc}
		\toprule
		Metric & Setting & $K_c^{\mathrm{tgt}}=3$ & $K_c^{\mathrm{tgt}}=5$ & $K_c^{\mathrm{tgt}}=7$ \\
		\midrule
		\multirow{3}{*}{depMSE}
		& Deployment transfer only & 0.0028 & 0.0028 & 0.0028 \\
		& Beam-head-only adaptation & 0.0028 & 0.0028 & 0.0028 \\
		& Full retraining & 0.0024 & 0.0028 & 0.0038 \\
		\midrule
		\multirow{2}{*}{Rate (bps/Hz)}
		& Beam-head-only adaptation & 5.5812 & 5.6500 & 5.0421 \\
		& Full retraining & 5.3661 & 5.7625 & 4.1293 \\
		\midrule
		\multirow{2}{*}{CRLB}
		& Beam-head-only adaptation & $1.16\times10^{-7}$ & $2.56\times10^{-6}$ & $6.06\times10^{-7}$ \\
		& Full retraining & $9.66\times10^{-8}$ & $2.95\times10^{-7}$ & $6.68\times10^{-7}$ \\
		\midrule
		\multirow{2}{*}{Trainable params}
		& Beam-head-only adaptation & 49,216 & 73,824 & 98,432 \\
		& Full retraining & 6,068,080 & 6,095,760 & 6,123,440 \\
		\midrule
		\multirow{2}{*}{Deployment drift}
		& Beam-head-only adaptation & $3.11\times10^{-11}$ & $4.05\times10^{-11}$ & $5.66\times10^{-11}$ \\
		& Full retraining & 0.0029 & 0.0030 & 0.0035 \\
		\midrule
		\multirow{2}{*}{Best epoch for rate}
		& Beam-head-only adaptation & 8 & 8 & 8 \\
		& Full retraining & 27 & 28 & 30 \\
		\bottomrule
	\end{tabular}
	\caption{Site-specific deployment transfer verification for SGNN\_LLM\_SH under changed communication user counts. The source task uses $K_c^{\mathrm{src}}=2$, while the target tasks use $K_c^{\mathrm{tgt}}\in\{3,5,7\}$.}
	\label{tab:site_specific_transfer_multi_kc}
\end{table*}

Finally, Table~\ref{tab:site_specific_transfer_multi_kc} evaluates the most distinctive property of the proposed framework, namely user-count transferability. In this experiment, a source model is first trained under $K_c^{\mathrm{src}}=2$, and then transferred to target tasks with changed communication user counts. The comparison is conducted between beam-head-only adaptation and full retraining.

The most important observation is that the deployment prediction remains highly stable after transfer. Specifically, the deployment transfer-only and beam-head-only adaptation settings both maintain a deployment MSE around $2.8\times 10^{-3}$ across all target user counts, while the deployment drift stays at the level of $10^{-11}$, which is essentially zero. In contrast, full retraining changes the deployment much more noticeably, as reflected by the much larger deployment drift values on the order of $10^{-3}$.

At the same time, beam-head-only adaptation is highly parameter-efficient. Across the three target settings, it requires only $49{,}216$ to $98{,}432$ trainable parameters, whereas full retraining always requires more than $6$ million trainable parameters. Despite this dramatic reduction in trainable parameters, beam-head-only adaptation still achieves competitive target-task performance. For example, when $K_c^{\mathrm{tgt}}=3$, beam-head-only adaptation even achieves a slightly higher rate than full retraining, while preserving the transferred deployment almost exactly.

Another important advantage is adaptation speed. Beam-head-only adaptation reaches its best rate within only $8$ epochs in all target settings, whereas full retraining requires $27$--$30$ epochs. This result strongly supports the claim that the learned deployment policy of SGNN\_LLM\_SH is largely site-specific and reusable across changed user configurations, while the main adaptation burden lies in the beamforming head rather than in the whole model.

Overall, the transfer results show that the proposed framework not only performs well on the source task, but also provides a practically meaningful mechanism for low-cost adaptation under changed communication user counts. This property is particularly desirable for segmented pinching antenna-assisted ISAC systems, where repeated full retraining under every new user configuration would be computationally expensive.
\section{Conclusion}\label{section:5}

This paper investigated the joint design of antenna deployment, segment-wise transmit/receive partitioning, and beamforming for segmented pinching antenna-assisted ISAC systems. To address the coupled communication--sensing optimization problem under variable user configurations, we proposed a general learning framework that combines a CSI-induced self-graph encoder with an LLM backbone using LoRA adaptation, followed by two task-specific output heads for deployment and beamforming prediction, respectively. The proposed design enables permutation-invariant scenario representation and supports unified prediction of deployment, partitioning, and communication/sensing beamforming variables.

Simulation results demonstrated that the proposed framework achieves a favorable tradeoff among communication rate, sensing accuracy, and convergence behavior compared with representative baselines. In particular, the graph-enhanced LLM model with split output heads exhibited strong robustness under communication-ratio variation and imperfect CSI. More importantly, the user-count transfer results showed that the learned deployment policy remains highly stable when the communication user count changes, while competitive target-task performance can be recovered by adapting only the beamforming head. This indicates that the proposed framework is able to capture a site-specific deployment representation and significantly reduce retraining cost compared with full model retraining.

Future work will extend the proposed framework to more challenging settings, including dynamic mobility, more severe CSI uncertainty, hardware-aware deployment constraints, and online adaptation for real-time SWAN-ISAC operation.

\appendix
\section{Derivation of the Position CRLB}\label{app:crlb}
\setcounter{equation}{0}
\renewcommand{\theequation}{A.\arabic{equation}}

In this appendix, we provide the derivation of the position Cram\'er--Rao lower bound (CRLB) adopted in the main text. Consider one sensing target located at
\begin{equation}
	\mathbf{p} = [x, y, z]^T \in \mathbb{R}^{3},
\end{equation}
and let the $n$th deployed pinching antenna be located at
\begin{equation}
	\bm{\psi}_n = [x_n, y_n, z_n]^T, \quad n=1,\ldots,N.
\end{equation}
Under the near-field spherical-wave model, the steering response from the target position $\mathbf{p}$ to the deployed aperture is written as
\begin{equation}
	\mathbf{a}(\mathbf{p})
	=
	\left[
	\frac{\alpha e^{-j\frac{2\pi}{\lambda} d_1}}{d_1},
	\frac{\alpha e^{-j\frac{2\pi}{\lambda} d_2}}{d_2},
	\ldots,
	\frac{\alpha e^{-j\frac{2\pi}{\lambda} d_N}}{d_N}
	\right]^T,
\end{equation}
where
\begin{equation}
	d_n \triangleq \|\mathbf{p}-\bm{\psi}_n\|_2
	=
	\sqrt{(x-x_n)^2+(y-y_n)^2+(z-z_n)^2}.
\end{equation}

Let $\mathbf{u}\in\mathbb{C}^{N\times 1}$ denote the effective sensing transmit weight over the deployed transmit aperture. The received echo after matched processing is modeled as
\begin{equation}
	\mathbf{r}
	=
	\beta \, \mathbf{a}(\mathbf{p}) \mathbf{a}^T(\mathbf{p}) \mathbf{u}
	+
	\mathbf{n},
	\label{eq:appendix_echo_signal}
\end{equation}
where $\beta \in \mathbb{C}$ is the complex reflection coefficient of the target and $\mathbf{n}\sim\mathcal{CN}(\mathbf{0},\sigma_r^2 \mathbf{I})$ is complex Gaussian noise. For notational convenience, define
\begin{equation}
	s(\mathbf{p}) \triangleq \mathbf{a}^T(\mathbf{p})\mathbf{u},
\end{equation}
so that the mean of the received signal is
\begin{equation}
	\bm{\mu}(\mathbf{p})
	\triangleq
	\mathbb{E}[\mathbf{r}]
	=
	\beta \, \mathbf{a}(\mathbf{p}) s(\mathbf{p}).
\end{equation}

Our goal is to estimate the target position parameter vector
\begin{equation}
	\boldsymbol{\eta} \triangleq [x, y, z]^T.
\end{equation}
Since the observation in \eqref{eq:appendix_echo_signal} is complex Gaussian with mean $\bm{\mu}(\mathbf{p})$ and covariance $\sigma_r^2\mathbf{I}$, the Fisher information matrix (FIM) for $\boldsymbol{\eta}$ is given by
\begin{equation}
	\mathbf{J}(\boldsymbol{\eta})
	=
	\frac{2}{\sigma_r^2}
	\Re
	\left\{
	\left(\frac{\partial \bm{\mu}}{\partial \boldsymbol{\eta}}\right)^H
	\left(\frac{\partial \bm{\mu}}{\partial \boldsymbol{\eta}}\right)
	\right\},
	\label{eq:appendix_fim_general}
\end{equation}
where
\begin{equation}
	\frac{\partial \bm{\mu}}{\partial \boldsymbol{\eta}}
	=
	\left[
	\frac{\partial \bm{\mu}}{\partial x},
	\frac{\partial \bm{\mu}}{\partial y},
	\frac{\partial \bm{\mu}}{\partial z}
	\right].
\end{equation}

Therefore, the key step is to compute the derivatives of $\bm{\mu}(\mathbf{p})$ with respect to the target coordinates. By the product rule,
\begin{equation}
	\frac{\partial \bm{\mu}}{\partial \xi}
	=
	\beta
	\left(
	\frac{\partial \mathbf{a}(\mathbf{p})}{\partial \xi} s(\mathbf{p})
	+
	\mathbf{a}(\mathbf{p}) \frac{\partial s(\mathbf{p})}{\partial \xi}
	\right),
	\quad \xi\in\{x,y,z\}.
	\label{eq:appendix_mu_derivative}
\end{equation}
Since
\begin{equation}
	s(\mathbf{p})=\mathbf{a}^T(\mathbf{p})\mathbf{u},
\end{equation}
we have
\begin{equation}
	\frac{\partial s(\mathbf{p})}{\partial \xi}
	=
	\left(\frac{\partial \mathbf{a}(\mathbf{p})}{\partial \xi}\right)^T \mathbf{u}.
	\label{eq:appendix_s_derivative}
\end{equation}

Next, for the $n$th entry of $\mathbf{a}(\mathbf{p})$,
\begin{equation}
	a_n(\mathbf{p})
	=
	\frac{\alpha e^{-j\frac{2\pi}{\lambda} d_n}}{d_n},
\end{equation}
its derivative with respect to $\xi\in\{x,y,z\}$ is
\begin{equation}
	\frac{\partial a_n(\mathbf{p})}{\partial \xi}
	=
	\alpha e^{-j\frac{2\pi}{\lambda} d_n}
	\left(
	-\frac{1}{d_n^2}
	-
	j\frac{2\pi}{\lambda}\frac{1}{d_n}
	\right)
	\frac{\partial d_n}{\partial \xi},
	\label{eq:appendix_an_derivative_raw}
\end{equation}
where
\begin{equation}
	\frac{\partial d_n}{\partial x}=\frac{x-x_n}{d_n},\quad
	\frac{\partial d_n}{\partial y}=\frac{y-y_n}{d_n},\quad
	\frac{\partial d_n}{\partial z}=\frac{z-z_n}{d_n}.
	\label{eq:appendix_dn_derivative}
\end{equation}
Substituting \eqref{eq:appendix_dn_derivative} into \eqref{eq:appendix_an_derivative_raw}, we obtain
\begin{equation}
	\frac{\partial a_n(\mathbf{p})}{\partial \xi}
	=
	\alpha e^{-j\frac{2\pi}{\lambda} d_n}
	\left(
	-\frac{1}{d_n^2}
	-
	j\frac{2\pi}{\lambda}\frac{1}{d_n}
	\right)
	\frac{\xi-\xi_n}{d_n},
	\quad \xi\in\{x,y,z\}.
\end{equation}
Accordingly,
\begin{equation}
	\frac{\partial \mathbf{a}(\mathbf{p})}{\partial \xi}
	=
	\left[
	\frac{\partial a_1(\mathbf{p})}{\partial \xi},
	\ldots,
	\frac{\partial a_N(\mathbf{p})}{\partial \xi}
	\right]^T.
\end{equation}

Substituting \eqref{eq:appendix_s_derivative} into \eqref{eq:appendix_mu_derivative} yields
\begin{equation}
	\frac{\partial \bm{\mu}}{\partial \xi}
	=
	\beta
	\left[
	\frac{\partial \mathbf{a}(\mathbf{p})}{\partial \xi}\mathbf{a}^T(\mathbf{p})\mathbf{u}
	+
	\mathbf{a}(\mathbf{p})
	\left(\frac{\partial \mathbf{a}(\mathbf{p})}{\partial \xi}\right)^T \mathbf{u}
	\right].
	\label{eq:appendix_mu_derivative_final}
\end{equation}
Then, by substituting \eqref{eq:appendix_mu_derivative_final} into \eqref{eq:appendix_fim_general}, the $(i,j)$th entry of the FIM can be written as
\begin{equation}
	[\mathbf{J}(\boldsymbol{\eta})]_{i,j}
	=
	\frac{2}{\sigma_r^2}
	\Re
	\left\{
	\left(\frac{\partial \bm{\mu}}{\partial \eta_i}\right)^H
	\left(\frac{\partial \bm{\mu}}{\partial \eta_j}\right)
	\right\},
	\quad \eta_i,\eta_j \in \{x,y,z\}.
\end{equation}

Finally, the CRLB matrix for target position estimation is given by
\begin{equation}
	\mathbf{C}(\boldsymbol{\eta})
	=
	\mathbf{J}^{-1}(\boldsymbol{\eta}),
\end{equation}
and the scalar sensing metric adopted in the main text is defined as
\begin{equation}
	\mathrm{CRLB}
	=
	\mathrm{tr}\!\left(\mathbf{J}^{-1}(\boldsymbol{\eta})\right).
\end{equation}
A smaller value of $\mathrm{CRLB}$ indicates a more accurate target position estimate and hence better sensing performance.

The above derivation shows that the sensing CRLB depends explicitly on the deployed antenna positions, the segment-dependent effective sensing aperture, and the sensing beamforming vector. Therefore, in the considered SWAN-ISAC system, antenna deployment, segment-wise partitioning, and beamforming are intrinsically coupled through both the communication objective and the sensing FIM structure.

\bibliographystyle{IEEEtran}
\bibliography{ref}

\end{document}